# Comparison between CS and JPEG in terms of image compression

Marija Milinković, Danko Petrić
University of Montenegro,
Faculty of Electrical Engineering
Podgorica, Montenegro

*Abstract*—The comparison between two approaches, JPEG and Compressive Sensing, is done in the paper. The approaches are compared in terms of image compression. Comparison is done by measuring the image quality versus number of samples used for image recovering. Images are visually compared. Also, numerical quality value, PSNR, is calculated and compared for the two approaches. It is shown that images, recovered by using the Compressive Sensing approach, have higher PSNR values compared to the images under JPEG compression. Difference is larger in grayscale images with small number of details, like e.g. medical images (x-ray). The theory is supported by the experimental results.

I. INTRODUCTION

The Shannon-Nyquist sampling theorem tells us that signal can be reconstructed if sampling frequency is at least twice higher then maximal signal frequency [1]. In many applications, the Nyquist frequency can be so high that we end up with too many samples and must compress in order to store or transmit them. Therefore, compression appeared as a necessity in order to ease storing and transmission of the signals. On the other side, during the transmission, signal samples can be lost, or can be corrupted by noise and considered as missing. These lost or corrupted signal values may be recovered under certain conditions, using the Compressive Sensing (CS) approach [1]-[4]. Depending on the application, different reconstruction approaches are used [5]-[20]. In this paper the comparison is made between commonly used image compression algorithm, JPEG, and CS approach, having in mind that both discard certain signal coefficients.

Compression algorithms are based on two basic presumptions: the imperfection of human perception and the specific properties of signal in certain transform domain such as Discrete Fourier Transform (DFT), Discrete Wavelet Transform (DWT), Discrete Cosine Transform (DCT), [1] etc.

JPEG algorithm is commonly used for image compression [1], [21], [22]. By using the JPEG algorithm, a notable compression ratio can be achieved, keeping high image quality. The standard JPEG algorithm is based on the 8×8 image blocks and uses DCT transform.

In recent years, the CS approach is intensively studied. The goal of this approach is to overcome the limits of the Shannon-Nyquist sampling theorem by recovering information about the signal using small set of available signal coefficients. To apply the CS, a certain conditions have to be satisfied, that are: random distribution of the available samples and sparse signal representation in a certain transform domain. Sparse representation means that signal, in certain domain, has a small number of nonzero coefficients in comparison with the signal length.

At the beginning, the CS approach was used in computed tomography, but later, the application field has grown. Many signals can be considered as sparse in certain transform domain, which makes them suitable for CS. For example, ISAR images are sparse in the domain of two-dimensional DFT, while digital image can be considered as a sparse in the DCT domain. CS approach is nowadays used in biomedical and communication signal analysis, etc.

The paper is organized in five sections. After Introduction, in Sections II and III basic information about JPEG and Compressive Sensing are provided. Section IV includes results of comparison and finally, paper is finished with concluding remarks in Section V.

II. JPEG COMPRESSION

JPEG compression algorithm uses image blocks of 8×8 size. After image dividing, a DCT transform of current image block is performed. DCT is defined by [1]:

$$DCT(k_1,k_2) = \frac{C(k_1)}{2}\frac{C(k_2)}{2}\sum_{i=0}^{7}\sum_{j=0}^{7}a(i,j)\cos\left(\frac{(2i+1)k_1\pi}{16}\right)\sin\left(\frac{(2j+1)k_2\pi}{16}\right) \quad (1)$$

Where:

$$C(k_1) = \begin{cases} \frac{1}{\sqrt{2}}, & \text{for } k_1 = 0 \\ 1, & \text{for } k_1 > 0 \end{cases} \quad C(k_2) = \begin{cases} \frac{1}{\sqrt{2}}, & \text{for } k_2 = 0 \\ 1, & \text{for } k_2 > 0 \end{cases} \quad (2)$$

and *a(i,j)* denotes a pixel of the original image. Transform coefficients are quantized using the quantization matrix. This matrix is calculated based on the compression quality that we want to achieve. After quantization, i.e. dividing DCT matrix with quantization matrix, coefficient at position (0,0) is called DC component. The rest coefficients are the AC components. DC represents the mean value of all coefficients in block. Quantization matrix is calculated by using the following relation:

$$\mathbf{Q}_{QF} = round(\mathbf{Q}_{50} \cdot q) \quad (3)$$

Where:

$$q = \begin{cases} 2 - 0.02 QF & \text{for } QF \geq 50 \\ \dfrac{50}{QF} & \text{for } QF < 50 \end{cases} \quad (4)$$

*QF* represents quality factor and $\mathbf{Q}_{50}$ is quantization matrix that is experimentally obtained. Matrix $\mathbf{Q}_{50}$ is shown in Fig. 1.

| 16 | 11 | 10 | 16 | 24 | 40 | 51 | 61 |
|----|----|----|----|----|----|----|----|
| 12 | 12 | 14 | 19 | 26 | 58 | 60 | 55 |
| 14 | 13 | 16 | 24 | 40 | 57 | 69 | 56 |
| 14 | 17 | 22 | 29 | 51 | 87 | 80 | 62 |
| 18 | 22 | 37 | 56 | 68 | 109| 103| 77 |
| 24 | 35 | 55 | 64 | 81 | 104| 113| 92 |
| 49 | 64 | 78 | 87 | 103| 121| 120| 101|
| 72 | 92 | 95 | 98 | 112| 100| 103| 99 |

Figure 1: Coefficients of the quantization matrix $\mathbf{Q}_{50}$

Quantization is applied:

$$DCT_q(k_1, k_2) = round\left(\frac{DCT(k_1, k_2)}{Q(k_1, k_2)}\right) \quad (5)$$

Matrix is then reordered in zigzag manner. AC components are coded using the Huffman Coding Table. This part of algorithm is lossless.
Quantization will leave us with matrix that has many zeros and coefficients mostly in the low frequency area with DC component at (0,0). At the decoder side, the inverse 2D DCT transform is applied:

$$a(i,j) = \sum_{i=0}^{7}\sum_{j=0}^{7} \frac{C(k_1)}{2}\frac{C(k_2)}{2} DCT_{dq}(k_1,k_2) cos\left(\frac{(2i+1)k_1}{16}\right) sin\left(\frac{(2j+1)k_2}{16}\right) \quad (6)$$

Where $DCT_{dq}(k_1, k_2)$ is dequantized DCT matrix:

$$DCT_{dq}(k_1, k_2) = DCT_q(k_1, k_2) \cdot Q(k_1, k_2) \quad (7)$$

We get the reconstructed image with an error proportional to the quantization step [1].
The area of interest for JPEG compression in this paper is after lossy part of JPEG Encoder shown in Fig. 2. Entropy coding and later decoding was not necessary for comparing algorithms. That being said, we needed program for first three blocks of Fig. 2. Quantization was implemented the way it is explained in this section for different QFs. Then number of non-zero coefficients is calculated for every DCT block. With this we knew how many DCT coefficient in CS procedure to put to zero and we were ready to compare performances of CS and JPEG in terms of image compression for the same number of non-zero coefficients.

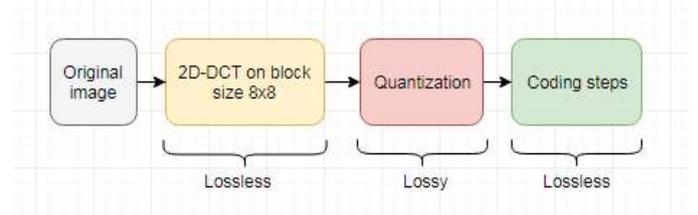

Figure 2. JPEG Encoder Block Schematic

### III. COMPRESSIVE SENSING

During the recent years of intensive research in the areas of CS, many different reconstruction approaches have been developed [1]-[20]. In order to be successfully reconstructed, signal needs to meet two important conditions. First, signal should have small number of nonzero samples in certain domain. This property is called sparsity. If a signal meets this condition then we can say that it has $K << N$ non-zero coefficients, where *K* is the number of signal components and *N* is the signal length. Second condition is the incoherence. It assumes that signal is rich of samples in original domain so that we can collect enough information about signal [1].

The CS procedure will be shortly described in the sequel. Using random measurement matrix $\mathbf{\Phi}(M, N)$, the set of random measurement are selected from the signal **f** of length *N*. Parameter *M* defines number of measurements taken in random manner. The measurement vector **y** can be defined as:

$$\mathbf{y} = \mathbf{\Phi f}. \quad (8)$$

Signal **f** has its counterpart in transform domain, vector **x**:

$$\mathbf{f} = \mathbf{\Psi x}, \quad (9)$$

where $\mathbf{\Psi}(N, N)$ is orthogonal basis matrix. Combining these two equations, we obtain the following relation:

$$\mathbf{y} = \mathbf{\Phi \Psi x} = \mathbf{A x}, \quad (10)$$

where **A** is the CS matrix of size $(M, N)$. The relation (10) is an undetermined system of equations. The aim is to solve *M* linear equations with *N* unknowns and it is done by using the optimization algorithms. There are many algorithms that are formed for this purpose, and all of them can be classified in the following categories: l1 minimization, greedy algorithms and total variation (TV) minimization.

In Fig. 3 [1], we can see illustration of CS concept where white fields represent zeros, i.e. missing coefficients.

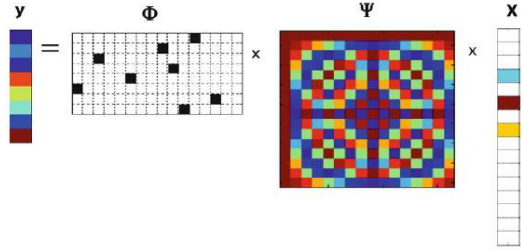

Figure 3: Illustration of the CS concept

For image reconstruction, the Total Variation (TV) minimization is used. Total Variation of image **x** represents the sum of the gradient magnitudes at each point and can be approximated as [1]:

$$TV(a) = \sum_{i,j} \|D_{i,j}x\|_2 \quad (11)$$

Where gradient for pixel at position $(i,j)$ is defined as:

$$D_{i,j}a = \begin{bmatrix} x(i+1,j) - x(i,j) \\ x(i,j+1) - x(i,j) \end{bmatrix} \quad (12)$$

Quality of reconstruction is measured by comparing PSNR(dB) of reconstructed images. PSNR is acronym of Peak to Signal to Noise Ratio. Mathematical definition is [3]:

$$PSNR = 10\log_{10}\frac{Q_{max}}{\sqrt{\frac{1}{MN}\sum_{i=0}^{M-1}\sum_{j=0}^{N-1}\left[x_{orig}(i,j) - x_{rec}(i,j)\right]^2}}, \quad (13)$$

where $x_{orig}$ and $x_{rec}$ are original and reconstructed images, $M$ and $N$ dimensions of **x** and $Q_{max}$ is maximal brightness of the image.

## IV. RESULTS

In this section we will present the results obtained by reconstructing images "lena.jpg" and "shepp-logan.png" using algorithms explained in previous sections. The goal is to compare the compression performance of two observed approaches, CS and JPEG.

The original images are shown in Fig. 4. The procedure consists of three steps:

1. JPEG compression with different compression ratio. Compression ratio (quality factor, QF) is changed from 10 to 90 with a step of 10. Calculation of the PSNR.

2. CS reconstruction using different number of missing image coefficients. It is important to note that we observed certain percentages of the missing samples. The number of missing samples is chosen to correspond to the number of zeros in the 2D DCT domain, left after the quantization in JPEG procedure. Calculation of the PSNR.

3. Comparison of the PSNRs, after JPEG compression and CS-based image reconstruction. The comparison is done for the same percentage of discarded samples in both approaches.

In Fig. 5 the values of PSNR(dB) for both observed images are shown.

In Fig. 8. the reconstructed/compressed images of shepp-logan.png are shown. Number of non-zero samples is 3.45, 7.46, 10.9 %, respectively. Fig. 9. shows the reconstruction/compression results of lena.jpg image. Number of non-zero samples is 5.52, 12.8, 18.4, 31.1 %, respectively, for both considered approaches – JPEG and CS.

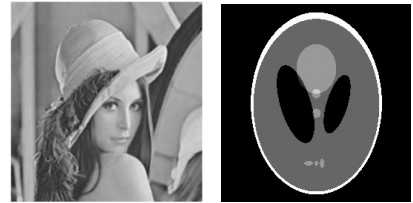

Figure 4. Original Images

TABLE I: PSNR VS NUMBER OF NONZERO SAMPLES USED IN JPEG AND CS APPROACHES

| QF | SAMPLES [%] | | PSNR[dB] JPEG | | PSNR[dB] CS | |
|---|---|---|---|---|---|---|
| | lena.jpg | Shepp-logan.png | lena.jpg | Shepp-logan.png | lena.jpg | Shepp-logan.png |
| 10 | 5.52 | 3.45 | 28.19 | 26.13 | 24.33 | 25.72 |
| 20 | 8.84 | 5.11 | 30.59 | 27.92 | 25.87 | 31.99 |
| 30 | 10.9 | 6.41 | 31.92 | 29.42 | 26.76 | 36.96 |
| 40 | 12.8 | 7.46 | 32.93 | 30.67 | 27.56 | 42.87 |
| 50 | 14.4 | 8.36 | 33.86 | 31.81 | 27.95 | 47.54 |
| 60 | 17.1 | 9.47 | 34.76 | 33.09 | 28.85 | 52.46 |
| 70 | 18.4 | 10.9 | 36.12 | 35.05 | 29.22 | 54.57 |
| 80 | 27.5 | 12.8 | 37.93 | 37.97 | 31.66 | 58.73 |
| 90 | 31.1 | 15.8 | 50.32 | 43.36 | 32.57 | 61.09 |

Figure 5. Lena and Shepp-Logan reconstruction PSNR(dB) (CS and JPEG)

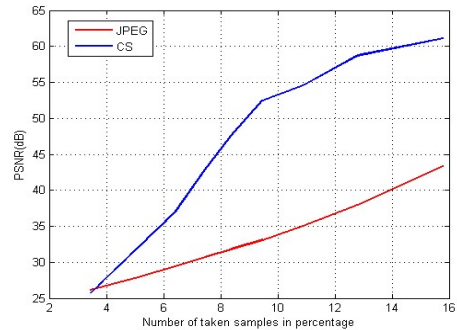

Figure 6. PSNR(dB) vs number of non-zero samples in Shepp-Logan reconstruction

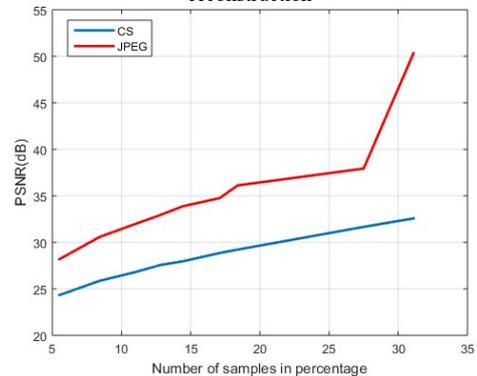

Figure 7. PSNR(dB) vs number of non-zero samples in Lena reconstruction

Based on the obtained results, the following can be concluded: when considering natural images as *lena.jpg*, a slightly better results are obtained by using the JPEG compression compared to the CS. This can be explained by the fact that, in JPEG, certain percent of low-frequency coefficients are used, which may not be the case in CS. CS randomly selects coefficients from all frequency plane. When considering the MRI images, that generally have less details compared to the natural images, the higher PSNR is obtained in CS compression.

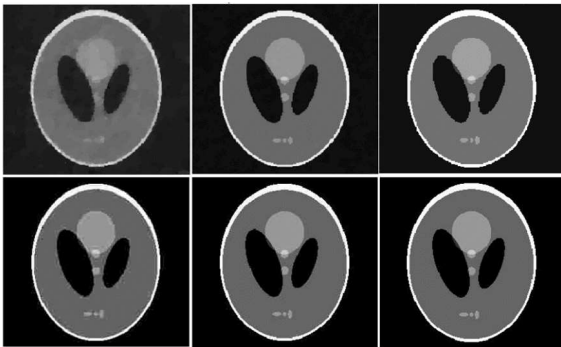

Figure 8. Reconstructed images "shepp-logan.png". First row – CS; Second row – JPEG;

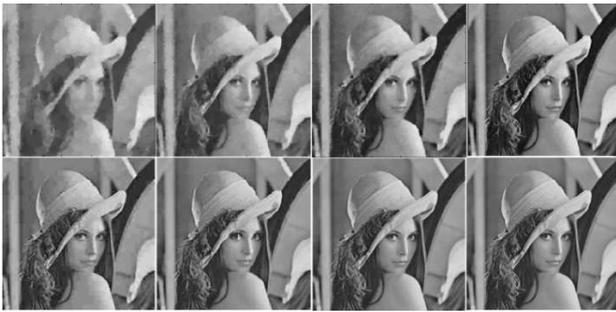

Figure 9. Reconstructed images "lena.jpg". First row – CS; Second row – JPEG;

## V. CONCLUSION

The comparison between JPEG and CS in terms of image compression is done in the paper. Percentage of missing samples in CS scenario corresponds to the number of nonzero coefficients, left in 2D DCT domain after the quantization with different quantization matrices in JPEG. The comparison is done in terms of PSNR value. The simpler images, such as MRI images, need less available (nonzero) samples to be reconstructed, compared to the natural images. This is expected because simpler images, in transform basis, have better sparsity than real images. Therefore, the CS is better solution when considering MRI images compression. However, the JPEG shows better performance in natural images due to the fact that natural images have more details and less sparsity in transform domain. Also, JPEG uses certain percent of low image coefficients from transform domain, that keep important information about the image.